\documentclass[aip,pof,amsmath,amssymb,
reprint,
author-year,
nofootinbib]{revtex4-1}
\usepackage[english]{babel}
\usepackage{stix}
\usepackage{graphicx}
\usepackage{xcolor}
\usepackage{bm}
\usepackage{ar}
\usepackage{subcaption}
\usepackage{relsize}
\usepackage{enumitem}
\usepackage{tikz}  
\usetikzlibrary{calc}
\usepackage{makecell}
\usepackage{hyperref}
\hypersetup{colorlinks=true,linkcolor=blue,citecolor=blue,linktocpage}

\definecolor{vf03}{HTML}{d0d1e6} 
\definecolor{vf05}{HTML}{3690c0}
\definecolor{vf10}{HTML}{045a8d}

\definecolor{ft03}{HTML}{fff7bc} 
\definecolor{ft05}{HTML}{ec7014}
\definecolor{ft10}{HTML}{993404}

\definecolor{olive}{HTML}{5da062}

\def\markup#1{#1}

\newcommand{\ftThrA}{$\bm \triangle$\hspace{-1.15em}{\color{ft03}$\triangle$}}
\newcommand{\ftFivA}{$\bm \lozenge $\hspace{-.8em}{\color{ft05}$\lozenge$}}
\newcommand{\ftTenA}{$\bm \pentagon$\hspace{-.9em}{\color{ft10}$\pentagon$}}
\newcommand{\ftThrB}{\textbf{---}\hspace{-1em}{\color{ft03}---}}
\newcommand{\ftFivB}{\textbf{---}\hspace{-1em}{\color{ft05}---}}
\newcommand{\ftTenB}{\textbf{---}\hspace{-1em}{\color{ft10}---}}

\makeatletter
\let\ORIbbl@fixname\bbl@fixname
\def\bbl@fixname#1{%
  \@ifundefined{languagealias@\expandafter\string#1}
    {\ORIbbl@fixname#1}
    {\edef\languagename{\@nameuse{languagealias@#1}}}%
}
\newcommand{\definelanguagealias}[2]{%
  \@namedef{languagealias@#1}{#2}%
}
\makeatother
\definelanguagealias{en}{english}

\begin{document}

\title{The effect of droplet coalescence on drag in turbulent channel flows}
\author{Ianto Cannon}
\affiliation{Complex Fluids and Flows Unit, Okinawa Institute of Science and Technology Graduate University, 1919-1 Tancha, Onna-son, Okinawa 904-0495, Japan}
\author{Daulet Izbassarov}
\affiliation{Department of Mechanical Engineering, Aalto University, FI-00076 Aalto, Finland}
\author{Outi Tammisola}
\affiliation{Linn\'{e} Flow Centre and SeRC (Swedish e-Science Research Centre), KTH Department of Engineering Mechanics, SE 100 44 Stockholm, Sweden}
\author{Luca Brandt}
\affiliation{Linn\'{e} Flow Centre and SeRC (Swedish e-Science Research Centre), KTH Department of Engineering Mechanics, SE 100 44 Stockholm, Sweden}
\author{Marco E. Rosti}
\email[Corresponding author: ]{marco.rosti@oist.jp}
\affiliation{Complex Fluids and Flows Unit, Okinawa Institute of Science and Technology Graduate University, 1919-1 Tancha, Onna-son, Okinawa 904-0495, Japan}
\date{\today}
\begin{abstract}
We study the effect of droplet coalescence on turbulent wall-bounded flows, by means of direct numerical simulations.  In particular, the volume-of-fluid and front-tracking methods are used to simulate turbulent channel flows containing coalescing and non-coalescing droplets, respectively.  We find that coalescing droplets have a negligible effect on the drag, whereas the non-coalescing ones steadily increase drag as the volume fraction of the dispersed phase increases: indeed,  at 10\% volume fraction, the non-coalescing droplets show a 30\% increase in drag, whereas the coalescing droplets show less than 4\% increase.  We explain this by looking at the wall-normal location of droplets in the channel and show that non-coalescing droplets enter the viscous sublayer, generating an interfacial shear stress which reduces the budget for viscous stress in the channel.  On the other hand, coalescing droplets migrate towards the bulk of the channel forming large aggregates,  which hardly affect the viscous shear stress while damping the Reynolds shear stress.  We prove this by relating the mean viscous shear stress integrated in the wall-normal direction to the centreline velocity.
\end{abstract}

\maketitle
\section{Introduction} \label{sec:introduction}
Two-fluid turbulent flows are found in many cases in industry and nature \citep{balachandar_turbulent_2010}, such as human arteries, industrial pipelines, and the injection of bubbles to enable drag reduction of ships \citep{ceccio_friction_2010}. In all of these cases, surfactants are known to have dramatic effects on the flow, often by preventing coalescence \citep{takagi_surfactant_2011}. However, due to the multi-scale nature of the problems, the mechanisms by which coalescence affects drag are not fully known and understood yet. Thus, the objective of this work is to explain how coalescence affects drag in wall-bounded flows.

Many experimental studies of surfactants in multiphase flow have been made. \citet{frumkin_surfactants_1947} were the first to describe the mechanism by which the rising speed of bubbles in water is reduced by surfactants (see\,\citep{levich_physicochemical_1962} for English version). \citet{descamps_airwater_2008} measured the wall shear stress in pipe flows of air bubbles in water,  and found that larger bubbles produced less drag. \citet{duineveld_bouncing_1997} studied pairs of bubbles rising in a vertical channel; he showed that coalescence is prevented when the surfactant concentration is above a critical value. As well as preventing coalescence, surfactants produce other effects on bubbles, such as clustering \citep{takagi_effects_2008}, reduction of rising velocity \citep{frumkin_surfactants_1947, levich_physicochemical_1962}, and reduction of shear-induced lift forces \citep{takagi_surfactant_2011}. Since all these effects can happen at the same time, the effect of different coalescence rate is difficult to highlight; on the other hand, simulations allow us to eliminate these effects, and focus solely on the impact of coalescence.

The majority of numerical multiphase flow studies have been made using interface-tracking methods, such as the front-tracking (FT) method \citep{unverdi_front-tracking_1992}. Front-tracking simulations of homogeneous-isotropic flows \citep{druzhinin_direct_1998} are well suited for investigating the effect of droplet size on the turbulent length scales,  such as bubble arrays \citep{esmaeeli_direct_1998,esmaeeli_direct_1999} or channel flows \citep{lu_dns_2006, dabiri_transition_2013,tryggvason_direct_2015, tryggvason_dnsassisted_2016, lu_effect_2017, ahmed_turbulent_2020}.  
An advantage of shear flow and channel-flow simulations is the ability to measure the effective viscosity and flow rate, which can then be compared with experiments. In the case of interface-tracking simulations of channel flows,  \citet{lu_dns_2006} simulated laminar bubbly upflows and downflows,  \citet{dabiri_transition_2013} showed that more deformable bubbles produced lower drag,  and \citet{lu_effect_2017} modelled bubbles with insoluble surfactant, and \citet{ahmed_turbulent_2020} with soluble surfactant, showing their main effects. However, none of the interface-tracking studies cited here includes a model for the breakup or coalescence of droplets, with only a few recent works tackling these phenomena\,\citep{lu_direct_2018,lu_multifluid_2019}.

Interface-capturing methods, such as the volume-of-fluid (VOF) method \citep{noh_slic_1976}, naturally allow coalescence and breakup of droplets \citep{elghobashi_direct_2019}. Interface-capturing simulations of homogenous isotropic turbulence \citep{dodd_interaction_2016, perlekar_droplet_2012, komrakova_numerical_2015, bolotnov_influence_2013} and shear flows \citep{de_vita_effect_2019, rosti_droplets_2019} have shed some light on the effect of coalescence on turbulence.  Notably, \citet{dodd_interaction_2016} and \citet{maxey_droplets_2017} showed that coalescence is a source of turbulent kinetic energy,  while breakup is a sink.  
\citet{scarbolo_coalescence_2015} investigated the effect of Weber number on breakup and coalescence, \citet{soligo_breakage_2019} modelled surfactant laden drops in turbulent channel flows, while \citet{bolotnov_detached_2011} used the level-set method to simulate bubbly channel flows. \citet{roccon_viscosity-modulated_2017} investigated the coalescence and breakup of large droplets in channel flow using the phase field method. \markup{Interface capturing methods are known to over-predict coalescence rates, because numerical coalescence occurs whenever the film thickness is less than the numerical grid spacing. In contrast, in real fluids film rupture occurs at molecular length-scales, which are in the tens of nanometres, orders of magnitude smaller than the Kolmogorov length\,\citep{tryggvason_multiscale_2013,soligo_breakage_2019}. A number of methods have been used to reduce the coalescence rate of interface capturing methods, such as adaptive grid refinement\,\citep{innocenti_direct_2021}, film drainage models\,\citep{thomas_multiscale_2010}, coupling to molecular dynamics simulations\,\citep{chen_large_2004}, and artificial forces\,\citep{de_vita_effect_2019}.}

In this paper, we use the front-tracking method to make simulations of droplets which cannot break up or coalesce, and we use the volume-of-fluid method to make simulations of droplets that easily break up and coalesce. \markup{As we are interested in the effects of coalescence, we do not use any methods to reduce the volume-of-fluid coalescence rate.} The two methods give idealized models of a mixture saturated with surfactants (FT), and completely clean mixture (VOF). Aside from coalescence and breakup, the physical properties (surface tension, viscosity, density, etc.) of the \markup{fluids} in the two methods are identical. To the authors' knowledge, this is the first direct comparison of coalescing and non-coalescing droplets in a turbulent channel flow.

The manuscript is organised as follows. First, in section~\ref{sec:method}, we describe the mathematical model governing the problem at hand and the numerical techniques used to solve them numerically. In particular, we describe our chosen interface-tracking and interface-capturing methods in more detail.  Section~\ref{sec:setup} reports the values of the parameters explored in our simulations. In section~\ref{sec:result}, we present statistics of the flow to elucidate how coalescence is affecting drag in the channel. Finally, section~\ref{sec:conclusion} gives conclusions and places them in the context of the current literature.

\section{Governing equations and flow geometry} \label{sec:method}

\begin{figure*}
	\begin{minipage}[c]{0.65\linewidth}
	\begin{tikzpicture}
		\node [above right,	inner sep=0] (image) at (0,0){
			\includegraphics[width=.9\textwidth]{bubblesFT}
		};
		\begin{scope}[x={($0.1*(image.south east)$)},y={($0.1*(image.north west)$)}]
			\node [above] at (-.2,8) {(a)};
			\node [] at (2.4,1.7){\text{\large$\boldsymbol{x}$}};
			\node [] at (.9,4.35){\text{\large$\boldsymbol{y}$}};
			\node [] at (0.35,1.5) {\text{\large$\boldsymbol{z}$}};
		\end{scope}
	\end{tikzpicture}	
	\begin{tikzpicture}
		\node [above right,	inner sep=0] (image) at (0,0){
			\includegraphics[width=.9\textwidth]{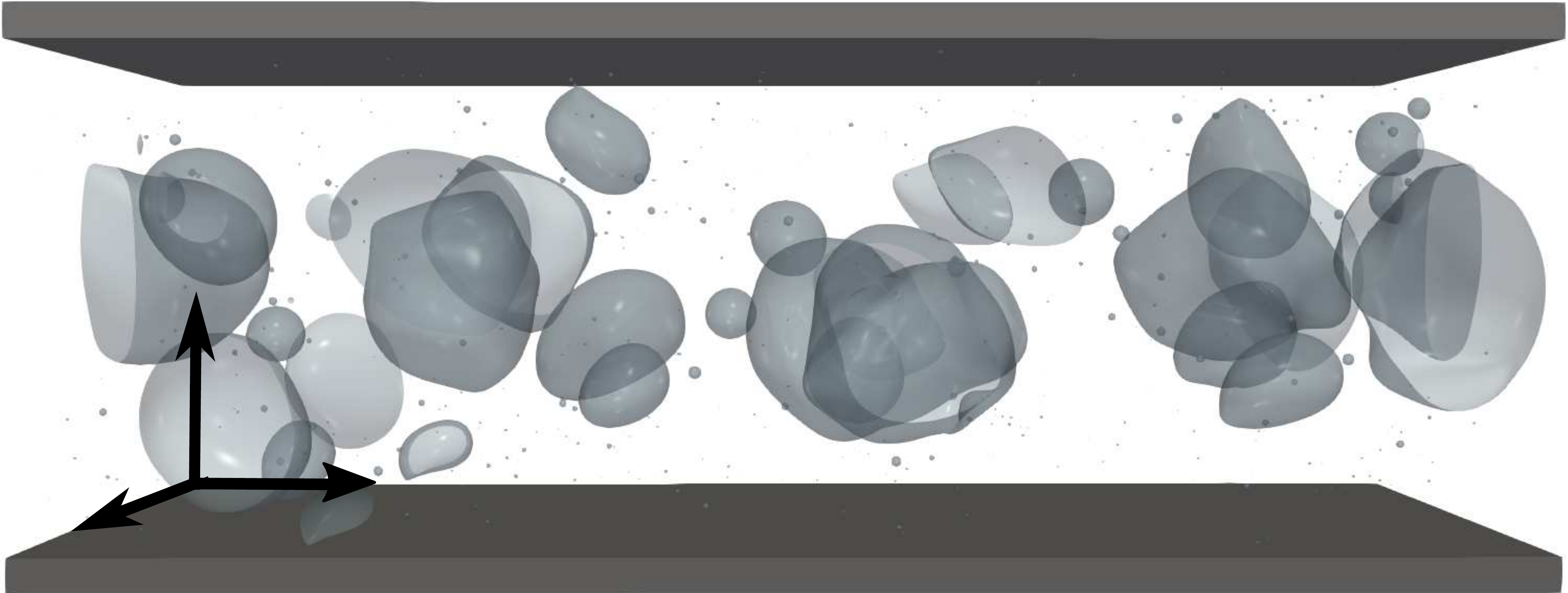}
		};
		\begin{scope}[x={($0.1*(image.south east)$)},y={($0.1*(image.north west)$)}]
			\node [] at (-.2,8) {(b)};
			\node [] at (2.5,1.9){\text{\large$\boldsymbol{x}$}};
			\node [] at (1.05,4.5){\text{\large$\boldsymbol{y}$}};
			\node [] at (0.45,1.5) {\text{\large$\boldsymbol{z}$}};
		\end{scope}
	\end{tikzpicture}	
\end{minipage}\hfill
\begin{minipage}[c]{0.35\textwidth}
	\caption{A snapshot of the simulation domain for a $10\%$ suspension of droplets simulated with \textbf{(a)} the front-tracking method (run FT10a in table~\ref{tab:runs}), and with \textbf{(b)} the volume-of-fluid method (run VOF10a). The orange and blue surfaces show the interface between fluid phases. The droplets in \textbf{(b)} can breakup and coalesce, giving rise to a range of sizes, whereas those in \textbf{(a)} cannot, thus remaining monodisperse.}
	\label{fig:bubblesFT_bubblesVOF}
\end{minipage}
\end{figure*}

We consider turbulent channel flows such as those shown in figure~\ref{fig:bubblesFT_bubblesVOF}. The numerical domain has size ${L_x\times L_y\times L_z=6L\times2L\times3L}$, where $L$ is the half-height of the channel. The flow is laden with an ensemble of $N$ droplets, initially spherical with radius $R=L/8$ and randomly arranged. We impose periodic boundary conditions in the streamwise ($x$) and spanwise ($z$) directions, while the non-slip and non-penetration boundary conditions are enforced at the two walls $y=0$ and $y=2L$. An imposed pressure gradient $G$, uniform throughout the domain and constant in time, sustains the flow in the $x$ direction. Balancing the forces on the fluid in the $x$ direction, we obtain an expression for the shear stress $\tau$ at the wall, ${\tau_w \equiv \langle \tau|_{y=0} \rangle_{xz} = G L}$, showing that $\tau_{w}$ remains constant in time. Note that, here and in the rest of the manuscript, we use angled brackets to represent an average over the subscripted directions. 

The Cartesian components of the fluid velocity field $(u_1,u_2,u_3)\equiv(u,v,w)$ are found by solving the incompressible multiphase Navier-Stokes equations at each location $\boldsymbol{x}$,
\begin{align}
	(\rho u_i)_{,t} + (\rho u_i u_j)_{,j} &= (\mu u_{i,j} + \mu u_{j,i} )_{,j} - p_{,i} + G \,\markup{\delta_{i1}} + \gamma \kappa n_i \, \markup{\delta_S(\boldsymbol{x})},\label{eq:NS}\\
	\label{eq:comp} u_{i,i} &= 0,
\end{align}
where $i,j \in \{1,2,3\}$. Throughout this article, we use Einstein notation\,\citep{einstein_grundlage_1916} where repeated indices are summed over, and the subscript comma denotes partial differentiation, i.e., ${\mathcal{F}_{,i} \equiv \frac{\partial \mathcal{F}}{\partial x_i}}$. The scalar $p$ is the pressure field used to enforce the incompressibility constraint stated in equation~(\ref{eq:comp}). The density $\rho$ and dynamic viscosity $\mu$ are the local weighted averages among the two phases, i.e., ${\rho = \phi\rho_d + (1-\phi)\rho_c}$ and ${\mu = \phi\mu_d + (1-\phi)\mu_c}$, where subscripts $_d$ and $_c$ denote properties of the dispersed and continuum phases respectively. In the above, $\phi$ represents the volume fraction of the dispersed phase in each computational cell of the domain, with $\phi=1$ in the dispersed phase and $\phi=0$ in the continuum phase. \markup{The Kronecker delta $\delta_{ij}$ is used to ensure that the pressure gradient is imposed in the $x$ direction.} The last term on the right hand side of equation (\ref{eq:NS}) is the \markup{volumetric formulation of the surface tension~\,\citep{popinet_numerical_2018}}; it is the product of the surface tension coefficient $\gamma$, the interface local curvature $\kappa$, and the unit normal to the interface ${n_i}$. Note that we used \markup{$\delta_S(\boldsymbol{x})$} in equation~(\ref{eq:NS}) to represent the \markup{surface delta function, which is zero everywhere except for the surface $S$ at the interface between the two phases. $\delta_S(\boldsymbol{x})$ has dimensions of inverse length.}

\subsection{Discretisation of the Navier-Stokes equations}
\label{sec:discretisation}
\markup{For simulations of coalescing and non-coalescing droplets, we use near-identical numerical methods to solve the momentum and continuity equations~(eqs.\@ \ref{eq:NS} and~\ref{eq:comp}). This ensures that any difference in our results is due to the droplets, not the integration scheme.}

\markup{Equations~\ref{eq:NS} and~\ref{eq:comp}} are solved numerically using a finite difference method on a fixed Eulerian grid with a staggered arrangement, i.e., fluid velocities are located on the cell faces and all other variables (pressure, density, viscosity, volume-of-fluid, etc.) are located at the cell centres.  All the spatial derivatives appearing in the equations are discretised with second-order central differences, except for the convective terms in the FT simulations where the QUICK scheme~\,\citep{leonard_stable_1979} is used instead. In the single-phase (SP) and VOF simulations, time integration is performed with the Adams-Bashforth method. In the FT simulations, time integration is performed with a predictor-corrector method, in which the first-order solution (Euler method) serves as a predictor which is then corrected by the trapezoidal rule\,\citep{tryggvason_front-tracking_2001, farooqi_communication_2019}. Both schemes are second-order in time.  Finally,  in regards to the pressure solver, the fractional step technique \citep{kim_application_1985} presented by \citet{dong_time-stepping_2012} and \citet{dodd_fast_2014} is adopted, allowing for direct solution of a constant-coefficient Poisson equation using an FFT-based solver, even in the presence of density differences among the carrier and dispersed phases. 

\subsection{Volume-of-fluid method}
We use the volume-of-fluid (VOF) method to simulate droplets undergoing topological changes, i.e., coalescence and break-up. This is an Eulerian-Eulerian technique in which the fluid phases are tracked using the local volume fraction scalar field $\phi$. Since~\citet{noh_slic_1976}, a number of variants of the VOF method have been developed \,\citep{youngs_time-dependent_1982, youngs_interface_1984, puckett_almgren_bell_marcus_rider_1997a, rider_reconstructing_1998, xiao_simple_2005, yokoi_efficient_2007}. Here we use the multi-dimensional tangent of hyperbola for interface capturing (MTHINC) method, developed by~\citet{ii_interface_2012}. In this method, we use a smooth hyperbolic tangent function to approximate the interface,
\begin{equation} 
	\label{eq:vofInd}
	H \left( X, Y, Z \right) = \tfrac{1}{2} + \tfrac{1}{2}\tanh \big( \beta \left( P \left( X, Y, Z \right) + d \right) \big),
\end{equation}
where $\beta$ is a parameter controlling the sharpness of the interface, and $d$ a normalisation parameter to enforce $\iiint H\,dX\,dY\,dZ = \phi$ in each cell. $P$ is a three-dimensional function in the cell, with the same normal and curvature as the interface. Normals are evaluated using the Youngs approach\,\citep{youngs_time-dependent_1982}, while the surface tension force appearing in the momentum equation~(\ref{eq:NS}) is computed using the continuum surface force (CSF) approach\,\citep{brackbill_continuum_1992}. \markup{See \citet{rosti_numerical_2019} for a detailed description of the volume-of-fluid code employed in this work, and in several other works\,\citep{rosti_droplets_2019,de_vita_effect_2019}. See\,\citet{ii_interface_2012} and\,\citet{de_vita_numerical_2020} for validations against numerical benchmarks and experiments}.

\subsection{Front-tracking method}
We use the front-tracking (FT) method to simulate droplets that can deform, but cannot break up or coalesce. This is an Eulerian-Lagrangian scheme in which the interface between the phases is tracked by a ``front'', composed of triangular elements.  The method was introduced by \citet{unverdi_front-tracking_1992}, with many refinements over the past 30 years\,\citep{tryggvason_front-tracking_2001, tryggvason_direct_2011}, including techniques to correct for errors in volume conservation of the phases\,\citep{takeuchi_volume_2020}. 
The surface tension force acting on the $L$th element is a volume integral of the surface tension force from equation~(\ref{eq:NS}),
\begin{equation}
\begin{split}
	\label{eq:elementForce}
	\boldsymbol{F_L} 
	&= \iiint_V \gamma \kappa \boldsymbol{n} \, \markup{\delta_{A_L}(\boldsymbol{x})} \, dV 
	= \iint_{A_L} \gamma \kappa \boldsymbol{n} \, d{A}\\
	&= \iint_{A_L} \gamma (\boldsymbol{n} \times \nabla) \times \boldsymbol{n} \,d{A} 
	= \oint_{s_L} \gamma \boldsymbol{t} \times \boldsymbol{n} \,d{s},
\end{split}
\end{equation}
where $A_L$ and $s_L$ are the area and perimeter of the $\markup{L}$th element and $\boldsymbol{t}$ is the tangent to the perimeter. The force is then interpolated onto the Eulerian grid by means of a conservative weighting function and used to update the fluid velocity, which in turn is used to update the position of the interface. As the interface evolves, the unstructured grid can greatly deform,  resulting in a non-uniform grid. Thus, periodical restructuring  of the Lagrangian grid is performed to maintain a nearly uniform size, comparable to the Eulerian grid size. \markup{See \citet{muradoglu_simulations_2014} for a detailed description and validation of the front-tracking code employed in this work, and used in several other works\,\citep{izbassarov_front-tracking_2015,lu_effect_2017,ahmed_turbulent_2020}. Extensive tests of the front tracking method are shown in \citet{tryggvason_front-tracking_2001}}.

\section{Setup}
\label{sec:setup}
\begin{table*}
	\begin{tabular}{|c|c|c|c||c|c|c|c||c|c|c|} 
		\hline
		Run& marker& method &coalescence&$\Phi$ (\%) &$Ca_{0}$ & $We_{\tau}$ & $N$ & $Re_b$ & $u^+_{cen}$ & $We_b$ \\ \hline
		SP0& \textbf{---}&N/A&N/A&0 &N/A&N/A&0&2836&18.38&N/A\\
		FT3a&\ftThrA&FT&No&2.5& 0.10 & 1.14 & 110 &2813&18.24&279.0\\
		FT3b&\ftThrB&FT&No&2.5& 0.05 & 0.57 & 110 &2661&17.19&124.8\\
		FT5a&\ftFivA&FT&No&5& 0.10 & 1.14 & 220 &2827&18.50&281.8\\
		FT5b&\ftFivB&FT&No&5& 0.05 & 0.57 & 220 &2602&16.93&119.4\\ 
		FT10a&\ftTenA&FT&No&10& 0.10 & 1.14 & 440 &2815&18.46&279.4\\ 
		FT10b&\ftTenB&FT&No&10& 0.05 & 0.57 & 440 &2524&16.54&112.3\\ 
		VOF3a&\color{vf03}$\bigblacktriangleup$&VOF&Yes&2.5&0.10 & 1.14 & 110 &2803&18.21&277.1\\
		VOF3b&\color{vf03}\textbf{---}&VOF&Yes&2.5& 0.05 & 0.57 & 110 &2818&18.15&140.0\\
		VOF5a&\color{vf05}$\mdlgblklozenge$&VOF&Yes&5& 0.10 & 1.14 & 220 &2764&18.26&269.4\\
		VOF5b&\color{vf05}\textbf{---}&VOF&Yes&5&0.05 & 0.57 & 220 &2778&18.07&136.1\\
		VOF10a&\color{vf10}$\pentagonblack$&VOF&Yes&10& 0.10 & 1.14 & 440 &2689&18.31&254.9\\
		VOF10b&\color{vf10}\textbf{---}&VOF&Yes&10&0.05 & 0.57 & 440 &2685&17.78&127.1\\
		\hline
	\end{tabular}
	\caption{Details of each turbulent channel flow simulation performed in the present study. The first column gives a unique name to each run for ease of reference, and the second describes the colours and markers that are used in the following figures.  Input variables are shown in the subsequent columns in the middle, and output statistics are shown in the three rightmost columns.}
	\label{tab:runs}
\end{table*}
Due to the different nature of the numerical schemes used to describe the presence of the interface,  the numerical domain is discretised on two different sets of grids, both verified to provide grid-independent results. The non-coalescing-droplet simulations use a uniform grid in the homogenous directions and a non-uniform grid in the wall-normal direction, with minimum spacing $\Delta Y_{FT} = 3L \times 10^{-3}$ at the channel wall. The minimum spacing in wall units is $\Delta Y^+_{FT} \equiv u_{\tau} \Delta Y_{FT} / \nu = 0.5$, where $ u_{\tau}$ and $\nu$ are defined later in this section. Overall, the grid size for the non-coalescing droplet simulations \markup{(FT)} is $N_x \times N_y \times N_z=576\times240\times288$, which is comparable to that used in \citet{dabiri_heat_2015}\markup{, and gives around 24 Eulerian grid points per droplet diameter. Due to periodic restructuring, we also have around 24 Lagrangian grid points per droplet diameter.} The single-phase and coalescing-droplet simulations \markup{(VOF)} use a cubic uniform grid with spacing $\Delta Y^+_{VOF} = 0.8$, and total size $N_x\times N_y \times N_z = 1296 \times 432 \times 648$. \markup{This grid has 108 points per initial droplet diameter.} We use more grid points in \markup{the VOF} simulations in order to accurately resolve breakup and coalescence events throughout the domain. 

The values of the non-dimensional parameters used in the simulations are shown in table~\ref{tab:runs}.  We consider a total volume fraction of the dispersed phase in the range ${0\% \leq \Phi \leq 10\%}$, with the continuum phase being denser and more viscous than the droplet phase, as the density ratio is fixed equal to $\rho_c/\rho_d=50$ and the dynamic viscosity to $\mu_c/\mu_d=50$ for all runs. Thus, the kinematic viscosity $\nu\equiv\mu/\rho$ has ratio $\nu_c/\nu_d=1$ for all runs. \markup{The problem approaches the density and viscosity ratios of air in water ($\rho_{water}/\rho_{air}\approx830,\, \mu_{water}/\mu_{air}\approx55$) while still being numerically tractable.} The friction Reynolds number $Re_{\tau} \equiv { u_{\tau} L / \nu}$ is set to 180 for all runs, where $u_{\tau} \equiv \sqrt{ \tau_{w} /\rho_c}$ is the friction velocity. We define the capillary number as $Ca_0 \equiv \mu_c u_0 / \gamma$ (where $u_0$ is the bulk velocity of the single-phase channel flow) for which two values are considered, $Ca_0 = 0.05$ and $0.10$. Based on these capillary numbers, the friction Weber number $We_{\tau} \equiv \rho_c u_{\tau}^2 L / \gamma$ assumes values smaller or larger than unity. Finally,  $N$ is the number of droplets at the start of the simulation, which are initially identical spheres in a random arrangement. 

The three rightmost columns in table~\ref{tab:runs} report three output statistics: the bulk Reynolds number, $Re_b\equiv u_{b} L / \nu$,  where $u_b \equiv \langle u \rangle_{xyz\markup{t}}$ is the bulk velocity; the bulk Weber number, ${We_{b} \equiv \rho_c u_{b}^2 L / \gamma}$; the centreline velocity in plus units $u^+_{cen} \equiv \langle u|_{y=L} \rangle_{xzt}/u_{\tau}$. In the next section, we present these and other statistics of the channel flows, and discuss their implications.

\section{Results} \label{sec:result}
\begin{figure*}
\begin{minipage}[c]{0.65\linewidth}
	\begin{tikzpicture}
		\node [above right,	inner sep=0] (image) at (0,0){
			\includegraphics[width=\textwidth]{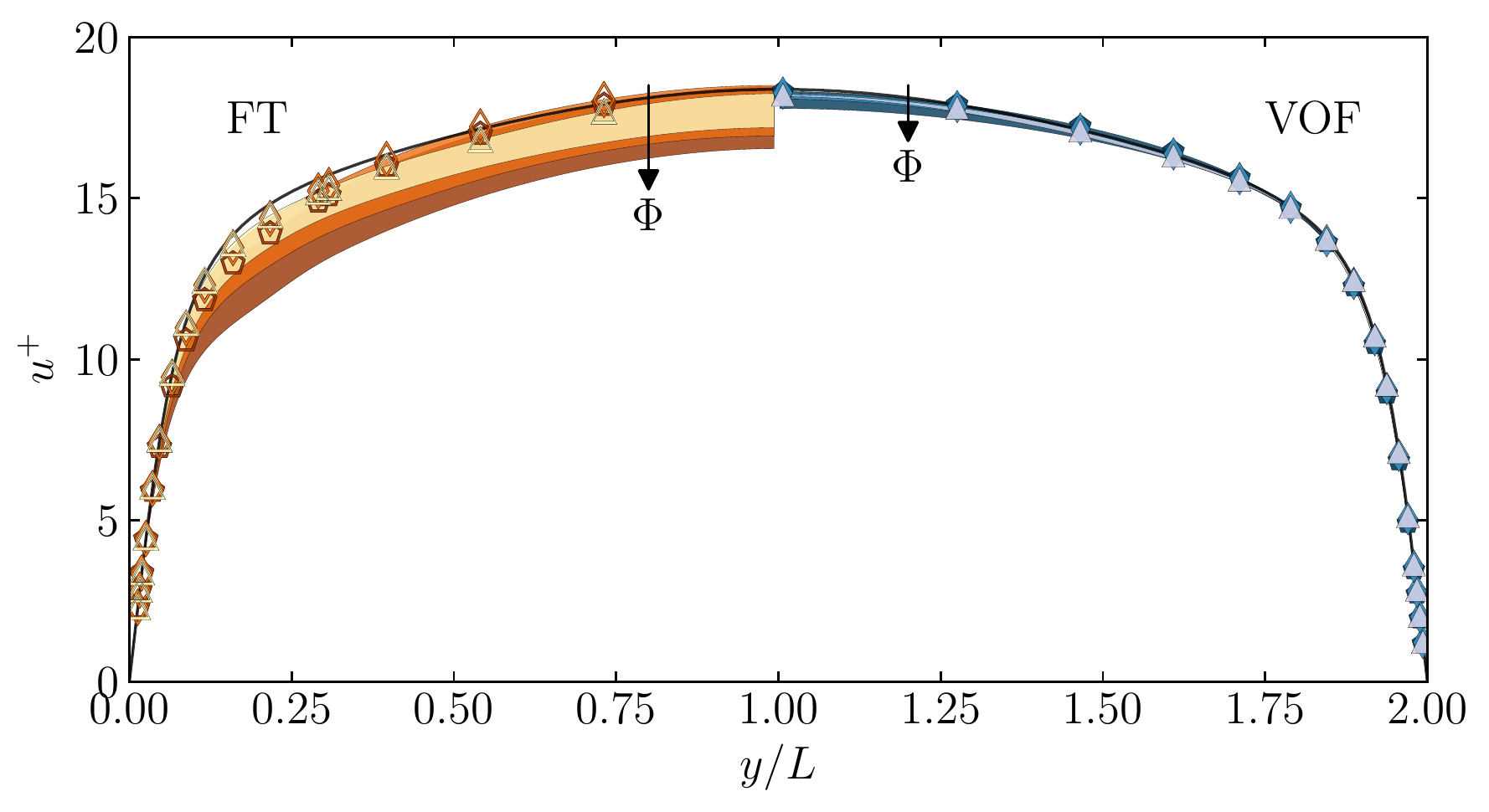}
		};
		\begin{scope}[x={($0.1*(image.south east)$)},y={($0.1*(image.north west)$)}]	
			\node[inner sep=0](inset) at (5.4,4.3) {
				\includegraphics[width=.5\textwidth]{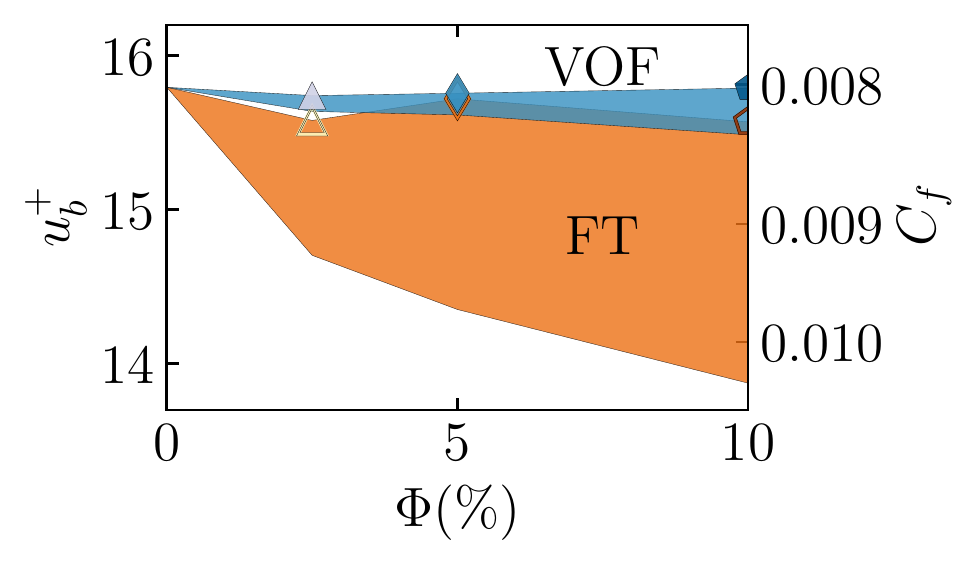}
			};
		\end{scope}	
\end{tikzpicture}
\end{minipage}\hfill
\begin{minipage}[c]{0.35\textwidth}
\caption{\textbf{Main:} Streamwise velocity profile in wall units $u^+$, against distance $y$ from the channel wall. The single-phase run (SP0) is shown as a black line. \markup{The profiles are symmetric about the centreline ($y=L$), so we have plotted runs with non-coalescing (FT), and coalescing (VOF) droplets on the left and right, respectively.} Each $Ca_0=0.1$ run is plotted using the marker listed in table~\ref{tab:runs}, while the regions between $u^+$ for $Ca_0=0.1$ and $Ca_0=0.05$ are shaded in colour. \textbf{Inset:} Dependence of bulk velocity $u_b^+$ on the total volume fraction of droplets $\Phi$. The skin-friction coefficient $C_f$ is shown on the right axis. Runs with coalescing droplets (VOF) are shown in \markup{blue, while runs with non-coalescing droplets (FT) are shown in orange}. Both plots show that drag increases with $\Phi$ and reduces with $Ca_0$ for all non-coalescing droplet runs, while very limited changes are observable for the coalescing droplet runs.}
	\label{fig:uVsY_uBVsPhi}
\end{minipage}
\end{figure*}
We consider turbulent channel flows in which droplets can coalesce, and compare the results with a configuration where coalescence is not allowed.  The flow is driven by a constant pressure drop, thus an increase in the flow rate or bulk velocity indicates drag reduction, while its reduction is evidence for drag increase.  We start by considering the profile of the streamwise velocity $u^+$ in the channel, reported in figure~\ref{fig:uVsY_uBVsPhi}. The single-phase run SP0 shows the typical velocity profile of a turbulent channel flow, with regions of high shear at the walls and a flattened profile in the channel centre. The runs with coalescing droplets (VOF) mostly collapse onto the single-phase profile, showing only a slight reduction in $u^+$ toward the centre. Whereas the runs with non-coalescing droplets (FT) show a significant reduction in $u^+$, which becomes more pronounced as $\Phi$ increases.  Also, in the coalescing droplets runs, variation of the capillary number produces little change in $u^+$, while in the non-coalescing runs, the change in $u^+$ with $Ca_0$ is much more substantial. 

This is quantified in the inset of figure~\ref{fig:uVsY_uBVsPhi}, which shows the bulk velocity in wall units $u^+_b \equiv \langle u \rangle_{xyzt} / u_{\tau}$ on the left axis, and the skin-friction coefficient ${C_f\equiv2\tau_w/\rho_c u_b^{2}}$ on the right axis. We see that, relative to the single-phase run, the coalescing droplets produce a maximum increase of 4\% in $C_f$, whereas the non-coalescing droplets produce a maximum increase of 30\%. In the case of non-coalescing droplets, the drag is highly dependent on $Ca_0$. The high $Ca_0$ (i.e., more deformable droplets) runs show little change in $C_f$ whereas the low $Ca_0$ (i.e., less deformable droplets) runs show a 30\% increase in $C_f$. Notably similar drag increases have been measured for rigid particles in channel flows by \citet{picano_turbulent_2015} and \citet{rosti_increase_2020}. Clearly, the coalescence of droplets in the channel has a profound effect on the flow. Throughout this section, we present additional statistics of the flows in order to shed light on the mechanisms of this effect.

\begin{figure*}
\begin{minipage}[c]{0.65\linewidth}
\begin{tikzpicture}
	\node [above right,	inner sep=0] (image) at (0,0){\includegraphics[width=.98\textwidth]{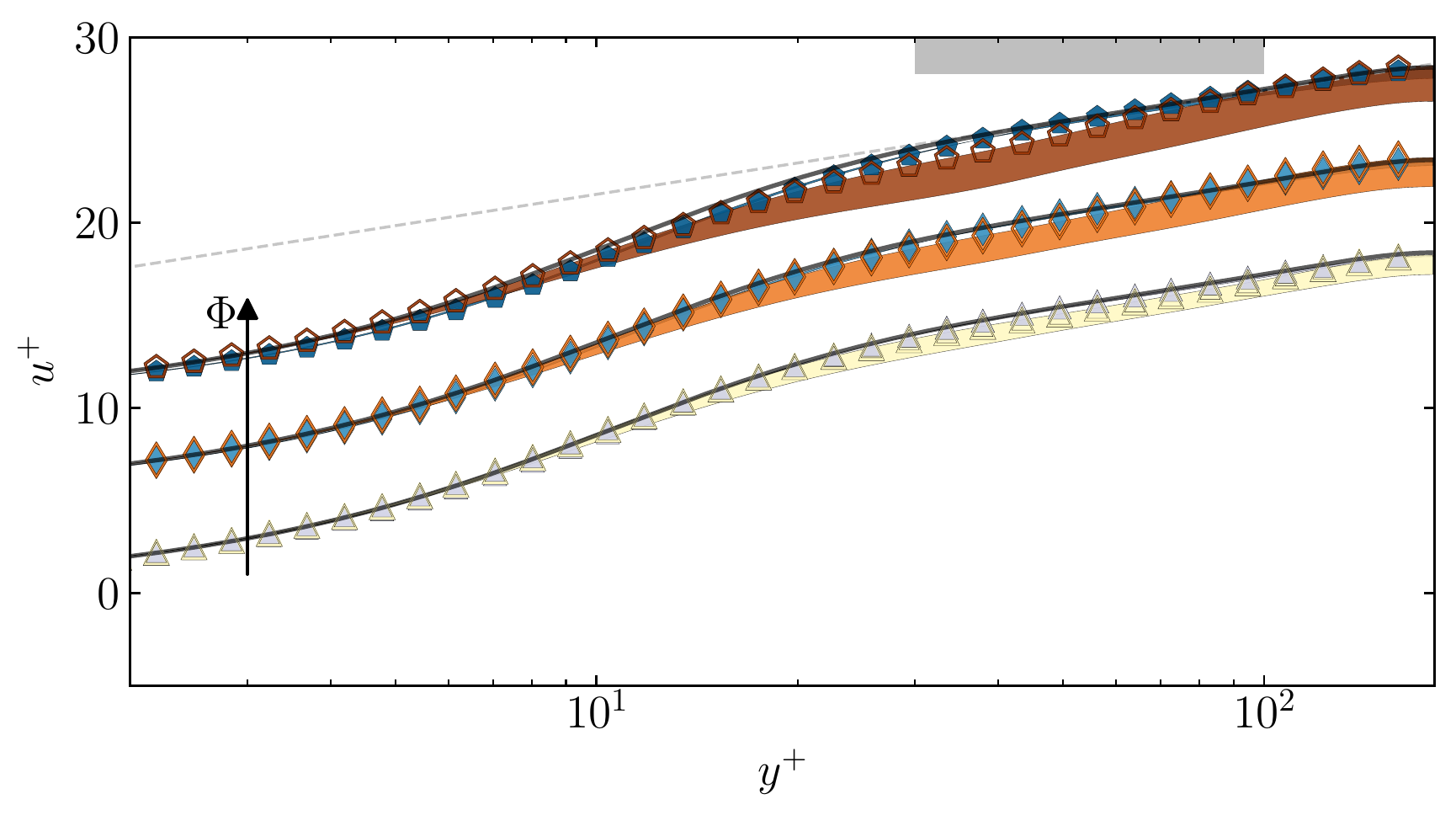}};
	\begin{scope}[x={($0.1*(image.south east)$)},y={($0.1*(image.north west)$)}]
		\node[] at (7.9,3.95) {\includegraphics[width=.37\textwidth]{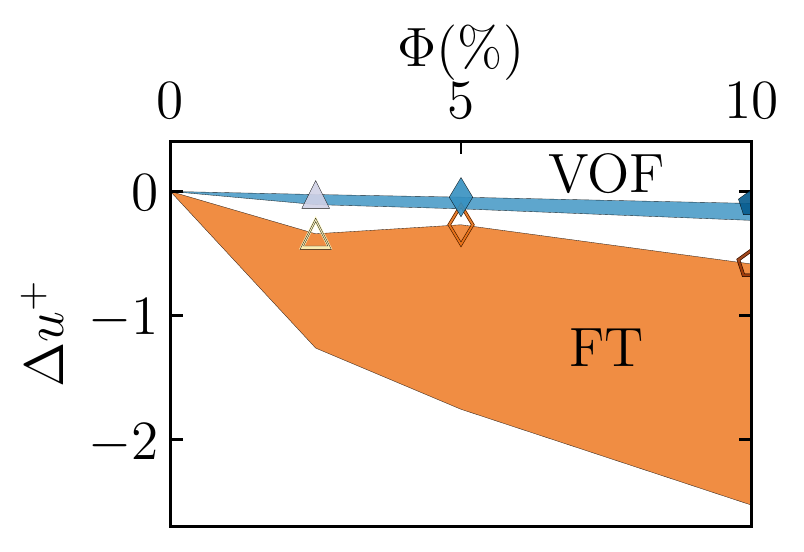}};
	\end{scope}
\end{tikzpicture}
\end{minipage}\hfill
\begin{minipage}[c]{0.35\textwidth}
\caption{\textbf{Main:} Velocity profiles in wall units $u^+$ and $y^+$. Each run is plotted using the marker listed in table~\ref{tab:runs}. For ease of comparison we have moved the $\Phi=5\%$ and $\Phi=10\%$ volume fraction profiles upwards by $u^+=5$ and $u^+=10$, respectively. In the region ${30 < y^+ < 100}$ shaded in grey, we fit a log-law equation ${u^+ = \frac{\ln y^+}{0.41} + 5.89 + \Delta u^+}$ (grey dashed line). \textbf{Inset:} The vertical shift $\Delta u^+$ for each run. Runs with coalescing droplets (VOF) are shown in \markup{blue, while runs with non-coalescing droplets (FT) are shown in orange}. Runs with coalescing droplets show only small shifts, whereas the runs with non-coalescing, less deformable droplets show significant drag increase.}
\label{fig:uPVsYP_uShiftVsPhi}
\end{minipage}
\end{figure*}
\markup{Figure~\ref{fig:uPVsYP_uShiftVsPhi} shows the velocity profile again, this time on a semi-log scale in wall units $u^+ \equiv u / u_{\tau}$, and $y^+\equiv y / \delta_\nu$, where $\delta_\nu\equiv \nu/u_\tau$ is the viscous lengthscale\,\citep{pope_turbulent_2000}. Away from the wall {and the channel centre $\delta_\nu << y << L$, i.e., the lengthscales affecting the flow are separated, and} the single phase flow profile is approximately parallel to a line with constant slope (the dashed line). This is a manifestation of the log-law for turbulent channel flows\,\citep{von_karman_mechanische_1930}, which can be derived by assuming the quantity $y^+\frac{du^+}{dy^+}$ has no dependence on $y^+$ or $y/L$ (complete similarity). The flow profiles with coalescing droplets in figure~\ref{fig:uPVsYP_uShiftVsPhi} are in excellent agreement with the log-law, suggesting that coalescing droplets do not break the scale separation. However, the flow profiles with non-coalescing droplets are not in such good agreement, because these droplets have constant size $R$, and $y \sim R$, so scale separation is prevented, hence $y^+\frac{du^+}{dy^+}$ shows a dependence on $y/R$.}

To further quantify the effect of coalescence on the flow, we fit a log-law function to each flow profile in the region ${30 < y^+ < 100}$. Our log law function has the form:
\begin{equation}
u^+ = \dfrac{\ln y^+}{0.41} + 5.89 + \Delta u^+,
\end{equation}
where $5.89$ is the $u^+$ intercept for run SP0, and $\Delta u^+$ is the shift relative to SP0. The inset of figure~\ref{fig:uPVsYP_uShiftVsPhi} shows how the vertical shift $\Delta u^+$ in the log-law region of the channel is affected by the volume fraction $\Phi$ and capillary number $Ca_0$ for the different cases.  Again, we see relatively small shifts for simulations with coalescing droplets, and large shifts for simulations with non-coalescing droplets. In particular,  $\Delta u^+$ grows in magnitude with $\Phi$, especially for the case with $Ca_0 = 0.05$. This reinforces our observations of the bulk streamwise velocity shown in the inset of figure~\ref{fig:uVsY_uBVsPhi}, that the less-deformable, non-coalescing droplets produce a significant drag increase.

\begin{figure*}
\begin{minipage}[c]{0.65\linewidth}
	\begin{tikzpicture}
		\node [above right,	inner sep=0] (image) at (0,0){
			\includegraphics[width=\linewidth]{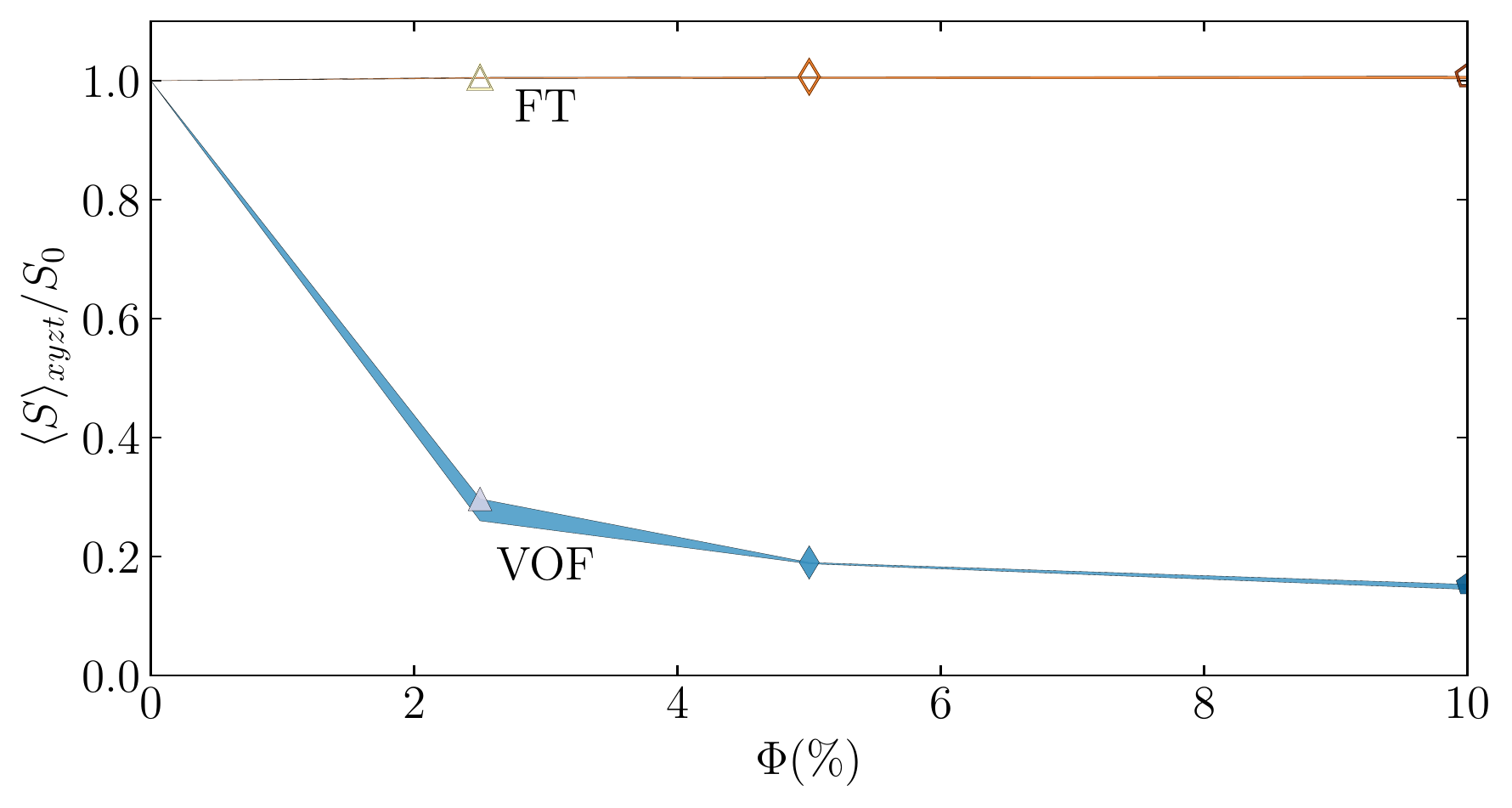}
		};
		\begin{scope}[x={($0.1*(image.south east)$)},y={($0.1*(image.north west)$)}]			
			\node[] at (7.3,5.9) {
				\includegraphics[width=.45\textwidth]{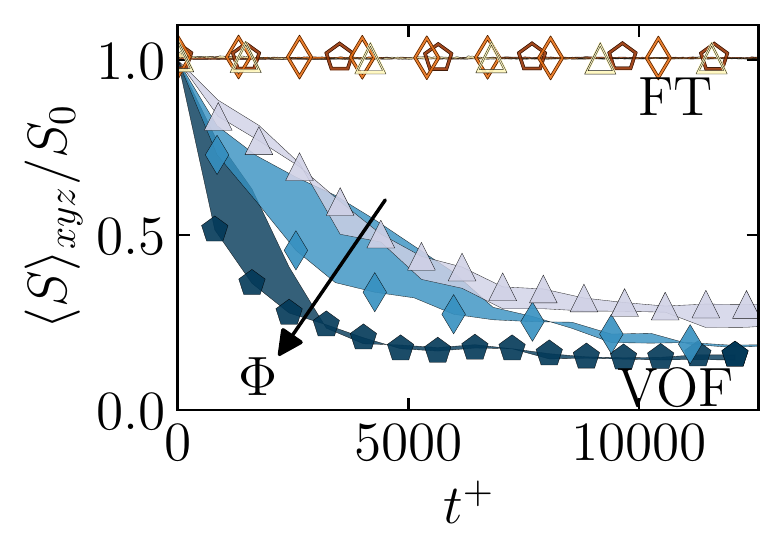}
			};
		\end{scope}
	\end{tikzpicture}
\end{minipage}\hfill
\begin{minipage}[c]{0.35\textwidth}
	\caption{\textbf{Main:} Dependence of the total interface area of the droplets $\langle S \rangle_{xyzt}$ on the total volume fraction $\Phi$. We have normalised each area by the total initial surface area $S_0$ of the droplets. The VOF runs \markup{(blue) show a major reduction in surface area due to coalescence, whereas the FT runs (orange)} show a slight increase, due to droplet deformation. \textbf{Inset:} Time history of the total interface area. Each run is plotted according to the colours and markers listed in table~\ref{tab:runs}. Note how the coalescing droplets (VOF) reach statistical equilibrium after $t^+\approx {8000}$, while the non-coalescing droplets (FT) converge very rapidly because of the absence of topological changes.}
	\label{fig:areaVsPhi_areaVsT}
\end{minipage}
\end{figure*}
To understand what generates the differences observed for configurations of coalescing and non-coalescing droplets, we focus our attention on the total surface area of the droplets. The total interface area is responsible for the overall surface tension stress, and impacts how droplets disperse across the channel. Figure~\ref{fig:areaVsPhi_areaVsT} shows how the total interface area at steady state $\langle S \rangle_{xyzt}$ depends on the total volume fraction $\Phi$ of the dispersed phase.  The figure shows that the non-coalescing droplets of the FT runs exhibit only 1\% increase in area, due to deformation from their initial spherical shape. On the other hand, the coalescing droplets of the VOF runs show more than 80\% reduction in interface area, as droplets coalesce and grow in size. In particular, when the volume fraction is large, droplets have a higher likelihood of colliding, and hence more coalescence, leading to a smaller value of ${\langle S \rangle_{xyzt} / S_0}$. 

\markup{For the coalescing droplets, the interface area ${\langle S \rangle_{xyzt} / S_0}$ shows no dependence on capillary number, differently from what was observed by \citet{lu_direct_2018} and \citet{rosti_droplets_2019}, who found that that as $Ca_0$ decreases, surface tension increases, the droplets become more stable to perturbations, hence larger, \markup{thus} leading to a smaller interface area ${\langle S \rangle_{xyzt} / S_0}$. However, in this case, $Ca_0<<1$, and the coalescing droplets are limited in size by the channel height, not surface tension. Figure~\ref{fig:bubblesFT_bubblesVOF}b supports this hypothesis, as the coalescing droplets are comparable in size to the channel height.}

The inset of figure~\ref{fig:areaVsPhi_areaVsT} reports the time history of the interface area: the cases with non-coalescing droplets (FT) rapidly converge to a statistically steady-state, whereas for the \markup{coalescing} droplets, convergence is reached long after, at about $t^+ \approx {8000}$. Interestingly, we observe that the coalescing droplet runs with larger capillary number (VOFa) converge to steady-state more rapidly than the smaller capillary number runs (VOFb), i.e., the \emph{larger} $Ca_0$ runs show a higher rate of coalescence, although the steady-state areas are roughly the same. This is in contrast with simulations of droplet coalescence in simple shear flow in laminar condition by \citet{shardt_simulations_2013}, which show droplet coalescence occurring only \emph{below} a critical $Ca_0$. However, as we shall discuss in the next paragraph, the $Ca_0=0.1$ droplets are more tightly confined in the channel centre than the $Ca_0=0.05$ droplets, thus leading to a higher rate of coalescence.

\begin{figure*}
\begin{minipage}[c]{0.65\linewidth}
	\includegraphics[width=\linewidth]{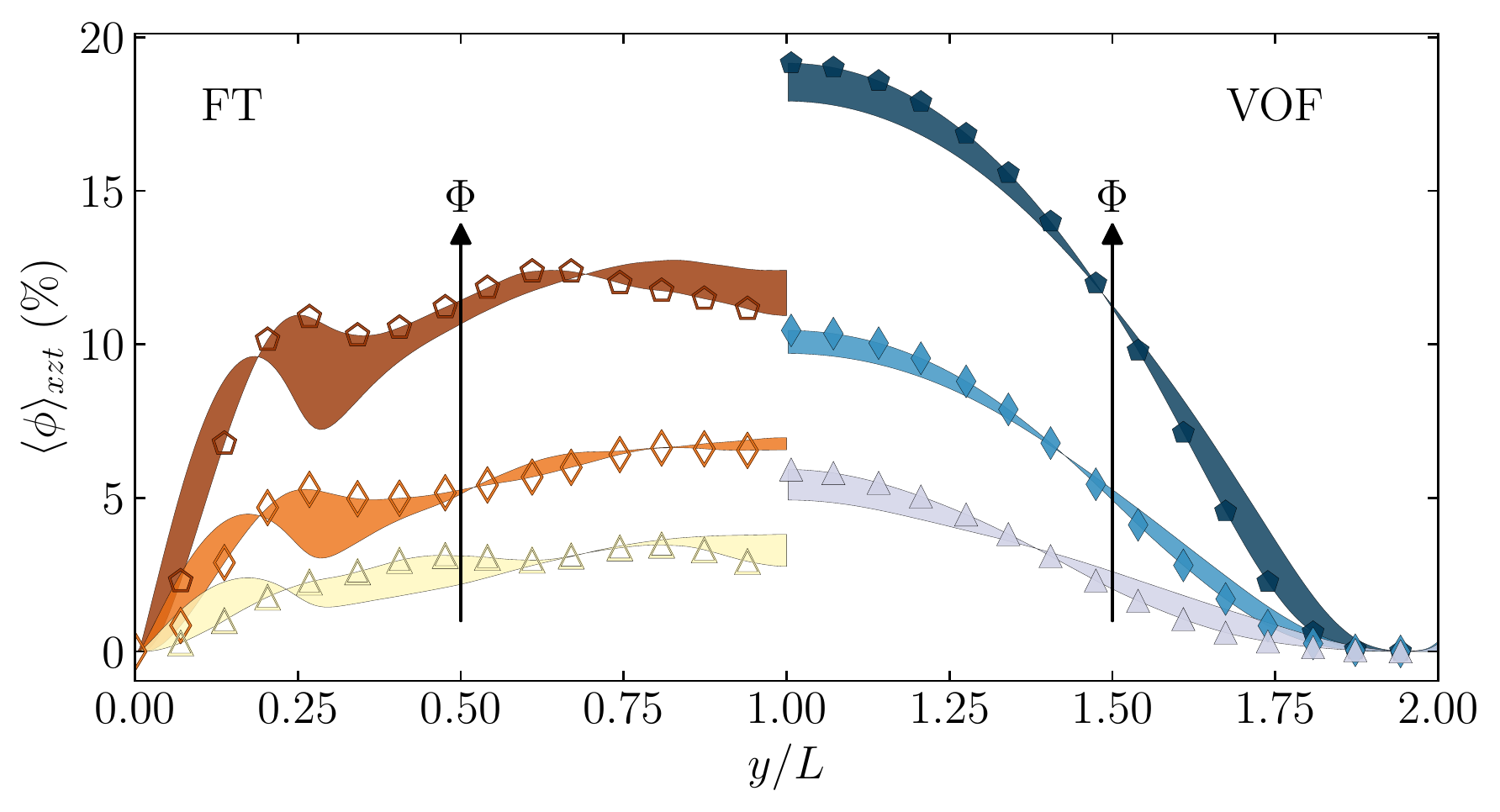}
\end{minipage}\hfill
\begin{minipage}[c]{0.35\textwidth}
\caption{Dependence of the mean volume fraction of droplets $\langle \phi \rangle_{xzt}$ on the distance $y$ from the channel wall. Each run is plotted using the colour and marker listed in table~\ref{tab:runs}. \markup{The profiles are symmetric about the centreline ($y=L$), so we have plotted runs with non-coalescing (FT), and coalescing (VOF) droplets on the left and right, respectively.} Note that for the runs with coalescing droplets, $\langle \phi \rangle_{xzt}$ peaks in the channel centre, whereas for the non-coalescing droplet runs, $\langle \phi \rangle_{xzt}$ shows a peak near the wall.}
\label{fig:PhiVsY}
\end{minipage}
\end{figure*}
Figure~\ref{fig:PhiVsY} shows how the volume fraction of the dispersed phase $\phi$ depends on the distance $y$ from the channel walls. The coalescing droplet profiles (VOF) clearly show a single peak at the channel centre, ${y=L}$: this peak arises as the droplets are driven toward the region of lowest shear ($y=L$) by a ``deformation-induced lift force''\,\citep{raffiee_elasto-inertial_2017,hadikhani_inertial_2018,alghalibi_inertial_2019}. Confinement in the channel centre leads to coalescence and the formation of large droplets, as seen in figure~\ref{fig:bubblesFT_bubblesVOF}b. 

The FT droplets cannot coalesce, and the droplet-droplet interaction produces a volume effect which forces them to spread across the channel: this manifests as an almost flat volume fraction in the region $0.5 L < y < L$ in figure~\ref{fig:PhiVsY}. Also, we see that the volume fraction tends to zero for $y<R=L/8$, as surface tension preserves the droplet radius $R$, and prevents the droplets from fully conforming with the flat channel walls. For all but one of the non-coalescing droplet runs plotted in figure~\ref{fig:PhiVsY}, $\langle \phi \rangle_{xzt}$ has a local maximum near the wall, in the region $0.15L<y<0.3L$. 
This phenomenon is \markup{due to the ``shear-gradient lift force'', which is known to act on particles in curved velocity profiles\,\citep{ho_inertial_1974,martel_inertial_2014,hadikhani_inertial_2018,alghalibi_inertial_2019}. Due to the curvature of the velocity profiles shown in figure~\ref{fig:uVsY_uBVsPhi}, the droplets experience different flow velocities on each side, resulting in a lift force toward the wall. From figure~\ref{fig:PhiVsY}, we also notice that the} more deformable droplets (FT3a, FT5a, and FT10a) produce a maximum which is further from the wall: this is mainly due to \textit{(i)} \markup{an increase in the deformation-induced} lift force, and to \textit{(ii)} a greater elongation of the droplets in the shear direction, producing a wider wall layer.

\begin{figure*}
\begin{minipage}[c]{0.65\linewidth}
	\includegraphics[width=\linewidth]{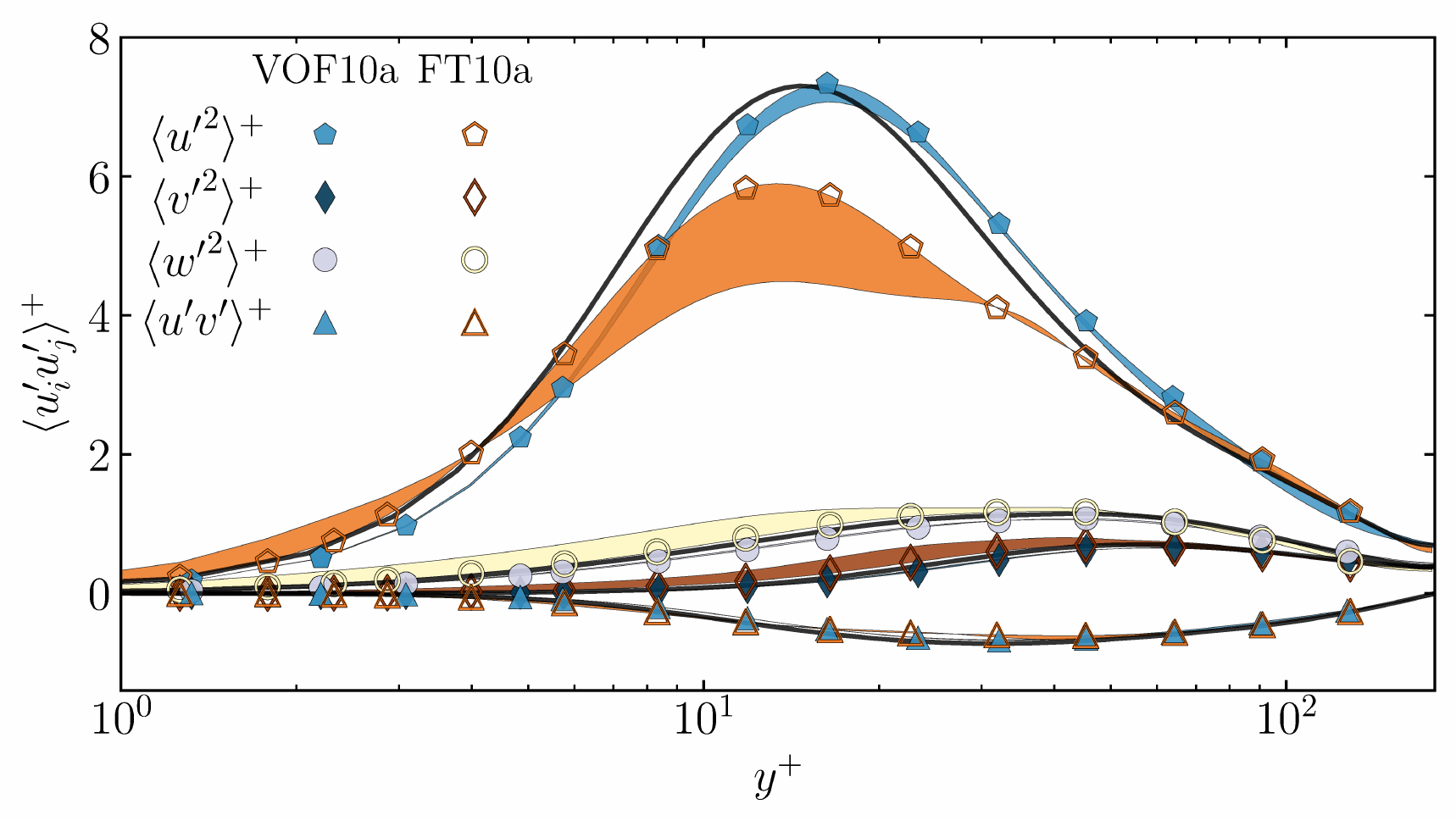}
\end{minipage}\hfill
\begin{minipage}[c]{0.35\textwidth}
	\caption{Variation of Reynolds stresses with the distance $y^+$ from the channel walls. Stresses for run SP0 are shown by solid black lines. For runs with droplets, the difference between the $Ca_0=0.1$ and $Ca_0=0.05$ stress is shaded in colour. The Reynolds stress components exhibit higher isotropy in the non-coalescing droplet runs (FT) than in the coalescing runs (VOF).}
	\label{fig:ReStressVsY}
\end{minipage}
\end{figure*}
We are now ready to investigate how droplets affect the turbulent flow, and we start by analysing the second-order statistics of the flow, which tell us how momentum is transferred across different parts of the channel.  Figure~\ref{fig:ReStressVsY} shows four of the six unique components of the Reynolds stress tensor in wall units $\langle u_i' u_j' \rangle^+ \equiv \langle u_i' u_j'\rangle_{xzt} / u_{\tau}^2$, with the single-phase (SP0) Reynolds stresses shown in black as reference. The coalescing droplets simulations (VOF) show little change in stresses relative to single-phase flow. Going from single phase to the non-coalescing droplets however, we see a reduction in the streamwise velocity fluctuations $\langle u'^2 \rangle^+$, and an increase in the wall-normal $\langle v'^2 \rangle^+$ and spanwise $\langle w'^2 \rangle^+$ velocity fluctuations. This shows that the isotropy of the turbulent flow has increased due to the presence of non-coalescing droplets. A similar effect has been observed for particle-laden turbulent channel flows, see e.g.\ \citet{picano_turbulent_2015}, in which particles redistribute energy to a ``more isotropic state'',  inducing an overall drag increase growing with the volume fraction of the dispersed phase. We infer that non-coalescing droplets have a back-reaction on the flow comparable to that of rigid particles, producing an increase in isotropy which correlates with an increase in drag. On the other hand, coalescing droplets \markup{produce a weaker back reaction on the flow, which shows little change in isotropy or drag}.

When compared to the other components of the Reynolds stresses, the shear stress $\langle u'v' \rangle^+$ shows only a small change due to the presence of droplets. However, as we shall see next, this shear stress opposes the pressure gradient in the channel, producing a profound impact on the drag.
\begin{figure*}
\begin{minipage}[c]{0.65\linewidth}
	\begin{tikzpicture}
		\node [above right,	inner sep=0] (image) at (2,0){\includegraphics[width=.93\textwidth]{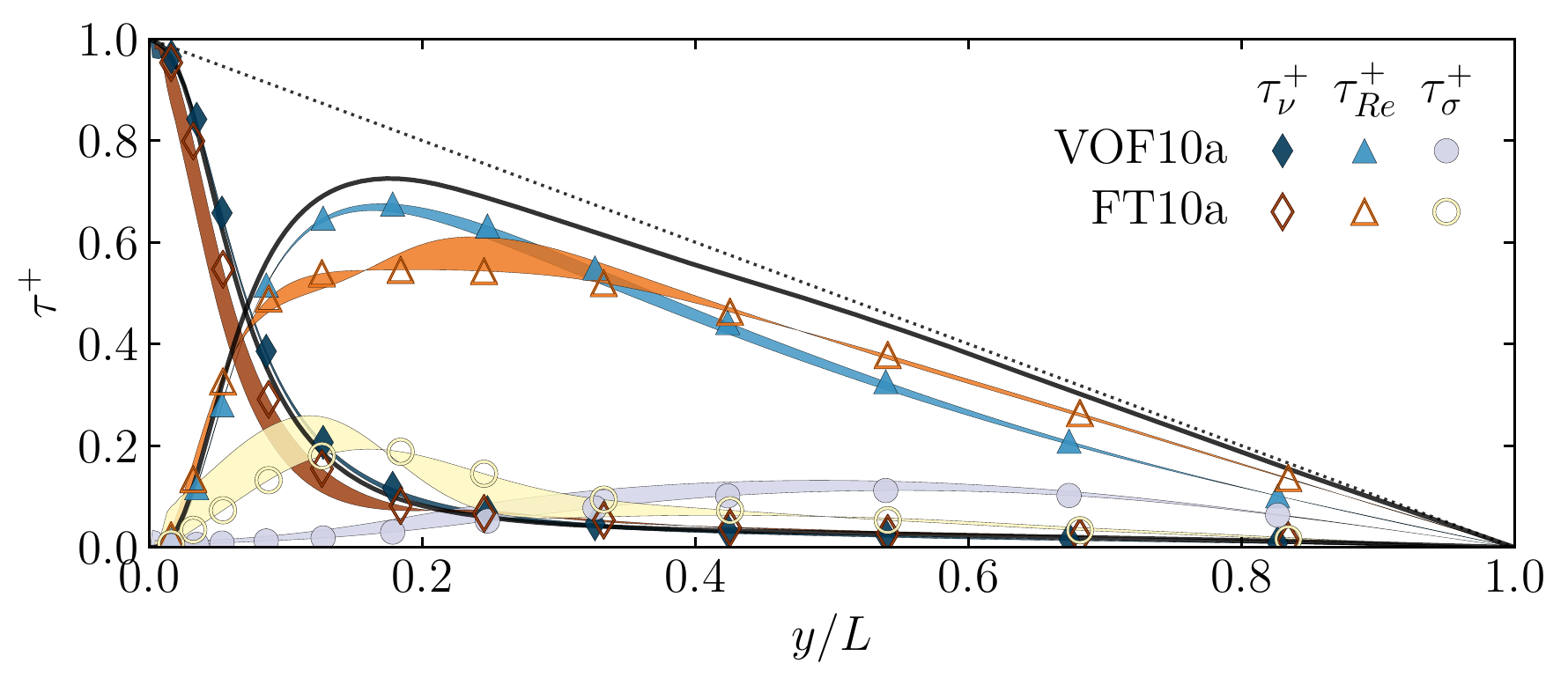}};	
		\begin{scope}[x={($0.1*(image.south east)$)},y={($0.1*(image.north west)$)}]
			\node [left] at (0,9) {(a)};
		\end{scope}
	\end{tikzpicture}	
	\begin{tikzpicture}
		\node [above right,	inner sep=0] (image) at (0,0){
			\includegraphics[width=.93\textwidth]{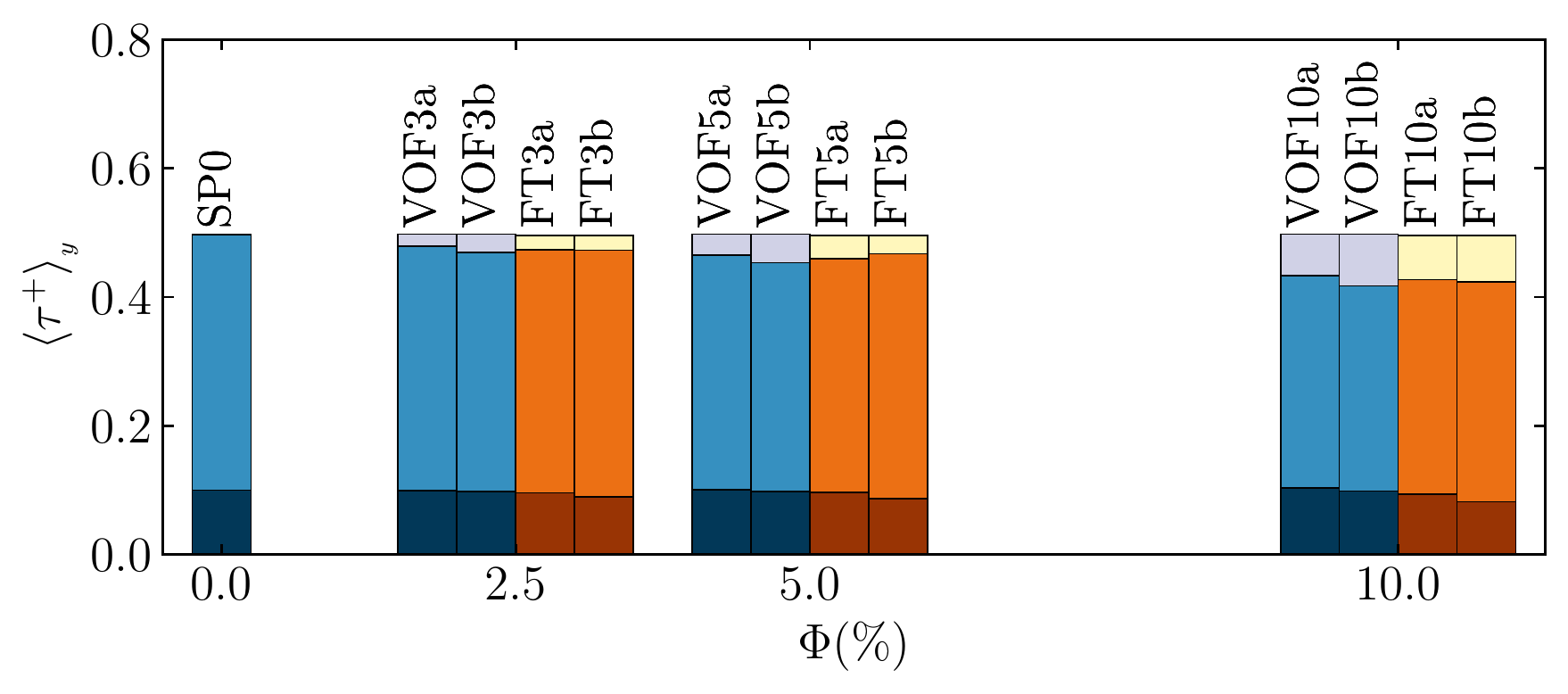}
		};
		\begin{scope}[x={($0.1*(image.south east)$)},y={($0.1*(image.north west)$)}]
			\node [left] at (0,9) {(b)};
		\end{scope}
	\end{tikzpicture}	
\end{minipage}\hfill
\begin{minipage}[c]{0.35\textwidth}
	\caption{\textbf{(a)} The balance of shear stresses as a function of the distance $y$ from the channel wall. The dashed line is the total stress budget. Stresses for run SP0 are shown by solid black lines. The differences between \markup{VOF10a and VOF10b stresses are shown in shades of blue, whereas the differences between FT10a and FT10b stresses are shown in shades of orange.} We see that $\tau_{\sigma}^+$ peaks near the wall for the runs with non-coalescing droplets (FT), but is spread across the channel for the coalescing runs (VOF).  The different stress distributions across the channel ultimately lead to different values of drag for coalescing and non-coalescing droplets. \textbf{(b)} Mean shear stresses for all runs. The stacked bars are $\langle \tau_{\nu}^+\rangle_y $, $ \langle \tau_{Re}^+\rangle_y $, and $\langle \tau_{\sigma}^+\rangle_y $ from bottom to top.}
	\label{fig:TauVsY_TauVsPhi}
\end{minipage}
\end{figure*}
The full shear stress balance for the multiphase problem under investigation can be obtained as follows. We start by taking average of the streamwise ($i=1$) component of equation~\ref{eq:NS}.
\begin{equation}
\begin{split}
	\langle (\rho u_1)_{,t} + (\rho u_1 u_j)_{,j})\rangle_{\markup{xzt}}  = 
	&\langle(\rho\nu u_{1,j} + \rho\nu u_{j,1} )_{,j}\rangle_{\markup{xzt}} 
	- \langle p_{,1}\rangle_{\markup{xzt}}\\
  &+ \langle G\markup{\delta_{11}} \rangle_{\markup{xzt}}  
	+ \langle \gamma \kappa n_1 \, \markup{\delta_S(\boldsymbol{x})} \rangle_{\markup{xzt}}.
\end{split}
\end{equation}
In fully developed turbulent channel flows, most of these terms average to zero, and the equation simplifies to
\begin{equation}
	\langle \rho u'v' \rangle_{\markup{xzt},y} = \nu\langle \rho u_{,y} \rangle_{\markup{xzt},y} + G + \langle \gamma \kappa n_1 \, \markup{\delta_S(\boldsymbol{x})} \rangle_{\markup{xzt}},
\end{equation}
where we have moved from the index notation $(u_1,u_2,u_3)$ to $(u,v,w)$ for the sake of clarity. \markup{Hereafter, for brevity we omit the subscripts $xzt$ on angled brackets}. Integrating from the wall $y=0$, to $y=\xi$ we obtain
\begin{equation}
	- G\xi = \Big[\langle \mu u_{,y}\rangle_{}
	- \langle \rho  u'v'\rangle_{}  \Big]^{y=\xi}_{y=0} 
	+ \int_{0}^{\xi} \langle \gamma \kappa n_1 \, \markup{\delta_S(\boldsymbol{x})} \rangle_{}  dy.
\end{equation}
The non-penetration boundary conditions at the walls enforce $v'=0$ and with $n_1=0$ at the wall, the lower limit of the right hand side is $ \langle \mu u_{,y}\rangle_{}|_{y=0} = \tau_w = GL$ by the definition of the wall shear stress. We relabel $y = \xi$ and obtain
\begin{equation}
	G(L-y) = \langle \mu u_{,y}\rangle_{}
	- \langle \rho u'v'\rangle_{}  
	+ \int_{0}^{y} \langle \gamma \kappa n_1 \, \markup{\delta_S(\boldsymbol{x})} \rangle_{}  dy.
\end{equation} 
By dividing the equation by $\tau_w$, we obtain the following dimensionless expression for the shear stress budget in the channel,
\begin{equation}
	1-{y}/{L} = \tau_{\nu}^+ + \tau_{Re}^+ + \tau_{\sigma}^+,
	\label{eq:stressBudget}
\end{equation}
where 
\begin{align}
	\tau_{\nu}^+ &\equiv \langle \mu u_{,y}\rangle_{}  / \tau_w, \label{eq:tauNu}\\
	\tau_{Re}^+ &\equiv - \langle \rho u'v'\rangle_{}  / \tau_w, \label{eq:tauRe}\\
	\tau_{\sigma}^+ &\equiv \int_{0}^{y}\langle \gamma \kappa n_1 \, \markup{\delta_S(\boldsymbol{x})} \rangle_{}  dy'/\tau_w,
\end{align}
are the viscous, Reynolds, and interfacial shear stresses, respectively. Here, we calculate the viscous stress and Reynolds stress using equations~\ref{eq:tauNu} and~\ref{eq:tauRe}, while the interfacial stress is calculated as the remaining part of the total budget in equation~\ref{eq:stressBudget}\footnote{Assuming the volume fraction $\phi$ is uncorrelated with the flow, we can separate the averages of the material properties and the flow velocity. To test our assumption, we measured the correlations $\nu \langle \rho u_{,y}\rangle_{}  - \nu \langle \rho \rangle_{}  \langle u\rangle_{,y}$ and $\langle \rho u'v'\rangle_{}  - \langle \rho \rangle_{}  \langle u'v'\rangle_{}$ for each of the FT runs, and found that the error in shear stress was always less than 3.5\% of $\tau_{w}$.}.

Figure~\ref{fig:TauVsY_TauVsPhi}a shows the balance of shear stresses from the channel wall ($y=0$) to the centre ($y=L$). In agreement with previous works\,\citep{pope_turbulent_2000}, the single-phase run (SP0) produces a viscous stress $\tau_{\nu}^+$ which is highest near the wall where the shear rate is maximum, and a Reynolds stress $\tau_{Re}^+$ which dominates for $y>0.1L$, where turbulent fluctuations abound.  

The coalescing droplet runs \markup{(VOF) in figure~\ref{fig:TauVsY_TauVsPhi}a} have an interfacial stress $\tau_{\sigma}^+$  which peaks around $y=0.5L$. This stress occurs due to \markup{the} droplet interfaces, which resist the deforming effects of turbulent fluctuations, at the detriment of the Reynolds stress. Note that $\tau_{\sigma}^+$ is larger for the smaller capillary number case (VOF10b compared to VOF10a), because the surface tension coefficient $\gamma$ is larger\markup{, so surface tension forces are larger}. 

The \markup{non-coalescing droplet runs (FT) in figure ~\ref{fig:TauVsY_TauVsPhi}a}, on the other hand, have very little interfacial stress $\tau_{\sigma}^+$ above $y>0.5$: \markup{instead}, the peak of $\tau_{\sigma}^+$ occurs at roughly the same wall-normal location $y$ as the peak in the volume fraction $\langle \phi \rangle_{xzt}$ seen in figure~\ref{fig:PhiVsY}. In both figure~\ref{fig:PhiVsY} and figure~\ref{fig:TauVsY_TauVsPhi}, the peak moves away from the wall when capillary number increases. A similar trend is \markup{also} observed for the location of the maximum turbulent kinetic energy production (not shown here). The correlation of $y$ locations for these three statistics suggests that the \markup{``wall layering'' and ``shear-gradient lift forces} discussed above, which produce a peak in $\langle \phi \rangle_{xzt}$ near the channel wall, are also responsible for $\tau_{\sigma}^+$ generation and kinetic energy generation.  The enhanced \markup{$\tau_{\sigma}^+$} close to the wall is compensated in the budget by a reduction in \markup{$\tau_{\nu}^+$} for the cases of non-coalescing droplets.

The averaged stresses are shown for all runs in figure~\ref{fig:TauVsY_TauVsPhi}b. The mean stresses are calculated by integrating $\tau_{\nu}^+$, $\tau_{Re}^+$, and $\tau_{\sigma}^+$ in the wall-normal direction $y$ from $0$ to \markup{$L$, for example,
\begin{equation}
\langle \tau_{\nu}^+ \rangle_y \equiv \frac{1}{L}\int_{0}^{L} \tau_{\nu}^+ \, dy.
\end{equation}
The averaged form of equation~\ref{eq:stressBudget} is $0.5 = \langle \tau_{\nu}^+ \rangle_y + \langle \tau_{Re}^+ \rangle_y + \langle \tau_{\sigma}^+ \rangle_y$, showing the averaged stresses are also in balance with the wall stress budget. We observe that for coalescing droplets, the dispersed fluid produces an interfacial stress $\langle \tau_{\nu}^+ \rangle_y$ which is compensated entirely by a reduction in Reynolds stress $\langle \tau_{Re}^+ \rangle_y$, with very little change in the viscous stress $\langle \tau_{\nu}^+ \rangle_y$. However, in the case of non-coalescing droplets the increase in interfacial stress $\langle \tau_{\sigma}^+ \rangle_y$ is compensated by a reduction in both the Reynolds stress $\langle \tau_{Re}^+ \rangle_y$, and the viscous stress $\langle \tau_{\nu}^+ \rangle_y$}. 

For the single-phase case, the dynamic viscosity \markup{$\mu$} is constant \markup{throughout the channel}, so the mean viscous stress is proportional to the centreline velocity,  
\begin{equation}
\begin{split}
	\markup{\langle \tau_{\nu}^+ \rangle_y 
	&= \frac{1}{L}\int_{0}^{L} \frac{\mu}{\tau_w} \frac{d \langle u \rangle_{xzt}}{dy} \,dy \\
	&= \frac{\mu}{L\tau_w} \Big[ \langle u\rangle_{xzt} \Big]_{y=0}^{y=L}\\
	&= \frac{\mu}{L\tau_w}u_{cen},}
	\label{eq:meanTauSig}
\end{split}
\end{equation}
and hence the variation of $\langle \tau_{\nu}^+ \rangle_y$ can be used to quantify drag in the channel, with a larger/smaller $\langle \tau_{\nu}^+ \rangle_y$ corresponding to drag reduction/increase.  For the multiphase problem, \markup{dynamic viscosity is different for the carrier phase and dispersed phases, and we should integrate $d\langle \mu u \rangle_{xzt}/dy$ to} the centreline, so the relationship between \markup{centreline velocity} and $\langle \tau_{\nu}^+ \rangle_y$ is not \markup{exactly linear}. \markup{However, due to the low volume fraction and low changes of viscosity,} we found that considering \markup{variation of the material properties ($\rho$, $\mu$) and variation of the fluid velocity as independent} produces only small changes in the averaged statistics. \markup{Hence} we can still relate the \markup{viscous} stress to the centreline velocity and thus to the drag changes in the multiphase simulations. Indeed, the three runs with the smallest bulk velocity $u_b^+$ in the inset of figure~\ref{fig:uVsY_uBVsPhi} are FT10b, FT5b, and FT3b, and the three runs with the smallest mean viscous stress $\langle \tau_{\nu}^+ \rangle_y$ are also FT10b, FT5b, and FT3b (figure~\ref{fig:TauVsY_TauVsPhi}b). Based on the above discussion, we can now relate the increased drag for non-coalescing droplets to the \emph{wall normal location} of the \markup{droplets}: the non-coalescing droplets in runs FT10b, FT5b, and FT3b encroach into the viscous wall region and oppose the shearing flow, reducing the viscous shear stress and thereby increasing drag.

\section{Conclusions} \label{sec:conclusion}
We perform direct numerical simulations of coalescing and non-coalescing droplets in turbulent channel flows to single out the effect of coalescence. Coalescing droplets are simulated using the volume-of-fluid method, and non-coalescing droplets with the front-tracking method.  We find that the droplets which are non-coalescing and less deformable produce an increase in drag, whereas the other droplets do not. We explained this by looking at the wall-normal location of droplets in the channel: the coalescing droplets experience a deformation-induced lift force, which drives them away from the shearing flow near the wall, out of the viscous sublayer; this is possible due to the coalescence which allows droplets to accumulate at the centreline. On the other hand, the non-coalescing droplets do not; indeed, non-coalescing droplets roughly behave as particles,  uniformly distributing across the channel, forming a wall layer and increasing the isotropy of the flow. In this case,  droplets remain in the viscous sublayer, generating an interfacial shear stress, which reduces the budget for viscous shear stress in the channel. From equation~\ref{eq:meanTauSig}, we relate a reduction in the viscous shear stress to a reduction in the centreline velocity, and ultimately to an increase in drag.

Our results agree well with the experiments carried out by \citet{descamps_airwater_2008}, who found that larger bubbles produce less drag; in our study, large droplets are obtained through coalescence,  and indeed produce less drag. Our proposed mechanism for drag increase is also similar to that proposed by \citet{dabiri_transition_2013}, who showed that less deformable bubbles enter the viscous sublayer, leading to an increase in viscous dissipation and an increase in drag. We offer two main developments. Firstly, we extend the study to coalescing droplets. Secondly, we believe that viscous shear stress is a better predictor of drag than viscous dissipation, as the proportionality between the mean viscous shear stress and centreline velocity (equation~\ref{eq:meanTauSig}) is exact for single-phase channel flows, and only slightly affected by the change in material properties. \markup{Although we made simulations at a density ratio of $\rho_c/\rho_d=50$, which is greater than that of oil in water ($\rho_{water}/\rho_{oil}\approx1.5$), but less than that of air in water ($\rho_{water}/\rho_{air}\approx830$),
comparison with experimental literature suggests that our current qualitative conclusions still hold for these flows.}

Our findings can help to better understand and control multiphase flows in a variety of applications, such as arteries, pipelines or ships. Through numerical experiments, we have been able to fully characterize the effect of coalescence alone, without the interference of other mechanisms which often arise in experiments with surfactants. How these results are affected by surfactant concentrations, will be the topic of future research.

\section*{Acknowledgements}
The authors acknowledge computer time provided by the Swedish National Infrastructure for Computing (SNIC), and by the Scientific Computing section of Research Support Division at OIST.  MER was supported by the JSPS KAKENHI Grant Number JP20K22402. OT was supported by Swedish Research Council grant VR 2017-0489

\section*{Data Availability Statement}
The data that support the findings of this study can be downloaded from \url{https://groups.oist.jp/cffu/cannon2021pof}

\end{document}